\documentclass[amsmat,amssymb,amsfonts,aps,prx,twocolumn,showpacs,superscriptaddress,showpacs]{revtex4-1}
\usepackage[english]{babel}
\usepackage{graphicx}
\usepackage{color}
\usepackage{dcolumn}
\usepackage{bm}        
\newcommand{\ang}{\mathrm{\AA}}

\begin{document}

\title{Optical response of metallic nanojunctions driven by single atom motion}

\author{F. Marchesin}
\affiliation{Centro de F\'isica de Materiales CSIC-UPV/EHU, Paseo Manuel de Lardizabal 4, 20018 Donostia-San Sebasti\'an, Spain}
\affiliation{Donostia International Physics Center (DIPC), Paseo Manuel de Lardizabal 5, 2001 Donostia-San Sebasti\'an, Spain}
\author{P.Koval}
\affiliation{Donostia International Physics Center (DIPC), Paseo Manuel de Lardizabal 5, 2001 Donostia-San Sebasti\'an, Spain}
\author{M. Barbry}
\affiliation{Centro de F\'isica de Materiales CSIC-UPV/EHU, Paseo Manuel de Lardizabal 4, 20018 Donostia-San Sebasti\'an, Spain}
\affiliation{Donostia International Physics Center (DIPC), Paseo Manuel de Lardizabal 5, 2001 Donostia-San Sebasti\'an, Spain}
\author{J. Aizpurua}
\email{aizpurua@ehu.eus} 
\affiliation{Centro de F\'isica de Materiales CSIC-UPV/EHU, Paseo Manuel de Lardizabal 4, 20018 Donostia-San Sebasti\'an, Spain}
\affiliation{Donostia International Physics Center (DIPC), Paseo Manuel de Lardizabal 5, 2001 Donostia-San Sebasti\'an, Spain}

\author{D. S\'anchez-Portal}
\email{sqbsapod@ehu.eus}
\affiliation{Centro de F\'isica de Materiales CSIC-UPV/EHU, Paseo Manuel de Lardizabal 4, 20018 Donostia-San Sebasti\'an, Spain}
\affiliation{Donostia International Physics Center (DIPC), Paseo Manuel de Lardizabal 5, 2001 Donostia-San Sebasti\'an, Spain}

\begin{abstract}
The correlation between transport properties across sub-nanometric metallic gaps and the optical response of the system is a complex effect which is determined by the fine atomic-scale details of the junction structure. 
As experimental advances are progressively accessing transport and optical characterization of smaller nanojunctions, a clear connection between the structural, electronic and optical properties in these nanocavities is needed. 
Using {\it ab initio}  calculations, we present here a study of the simultaneous 
 evolution of the structure and the  
optical response of a plasmonic junction 
 as the particles forming the cavity, two Na$_{380}$ clusters, approach and retract. 
Atomic reorganizations are responsible for a large hysteresis of the plasmonic response of the system, that
shows a jump-to-contact instability during the approach process and the formation of an atom-sized neck 
across the junction during retraction. 
Our calculations demonstrate that, due to the quantization of the conductance in metal nanocontacts, 
 atomic-scale reconfigurations play a crucial role in determining the 
optical response of the whole system.
We observe abrupt changes in the intensities and spectral positions
of the dominating plasmon resonances, and find a one-to-one correspondence between 
these jumps and those of the quantized transport as the neck cross-section diminishes. 
These results point out to an unforeseen connection between transport and optics at the atomic scale, which is 
at the frontier of current optoelectronics and can drive 
new options in optical engineering of signals driven by the motion and manipulation of single atoms. 
 
\end{abstract}


\maketitle

\section{Introduction}

Understanding and controlling the optical response of nanosystems is a key aspect in the search for nanoarchitectures that combine the current performance of miniaturized 
silicon-based electronics with the fast control and engineering of optical signal. 
A solution that combines the advantages of one and another inevitably needs 
to account for the difference in frequency and size ranges 
that characterize the electronic and optical regimes. 
One possibility to close the gap between electronics and photonics is provided by
the ability of light to 
coherently excite collective electronic excitations at the surface of 
metallic nanostructures,~\cite{Ozbay06,Brongersma10}
effectively bringing light to the nanometer-scale. These excitations, commonly 
referred to as surface plasmons,~\cite{Ritchie57,Barnes03,Pitarke07,Maier07,Plasmonics12} 
have the ability to localize and enhance light in the proximity of the surface of 
nanostructures,~\cite{Schuck05,AlonsoGonzalez12,Morton11,Barbry15,Steuwe15,Muhlschlegel15} 
thus establishing metal nanoparticles 
as relevant building blocks in current Nanophotonics.~\cite{Pelton08,Aberasturi15} 
Surface plasmons are central for the development and the high optimization level
of many techniques and processes such as 
field-enhanced 
spectroscopy,~\cite{Xu00,Bouhelier03,Neubrech08,Jensen08,Morton11,Sigle13,Aroca13,Shiohara14,Hakonen15}
thermotherapy,~\cite{Hirsch03,Huang08,Lal08}
photovoltaics,~\cite{Atwater10,Li12,Zhang13}
optical sensing,~\cite{GarciaEtxarri10,Polavarapua13,Virk14,Thacker14,Li15,Rodrigo15}
optical nanoengineering,~\cite{Oldenburg98,Fan10,Porta15}, 
or near-field 
microscopy.~\cite{Anderson00,Stoeckle00,Anger06,Huber09,Morton11,Bonnell12,Amenabar13,ZhangAizpurua13}

In this context, one option in the search for a proper electro-optical interface relies 
in pushing the limits of nanometric plasmonics down to the realm of the atomic-scale. 
It has recently been  shown that the interaction of metal 
surfaces in sub-nanometric proximity drives new optoelectronic phenomena, 
where an interplay between the photons, 
single electron transitions, 
plasmons, 
vibrations
and motion of atoms present in the junction, determines the complex outcome of the optical response including strong quantum effects and nonlinearities.~\cite{Dankwerts07,Marinica12}
On the one hand, strong non-local dynamical screening~\cite{Luo13} and 
quantum tunnelling~\cite{Zuloaga09,Esteban12,Savage12}
have been shown to drastically modify the optical response in a metallic sub-nanometric 
gap, establishing the limit of localization and enhancement of the optical fields 
far below the predictions from simple 
classical approaches.~\cite{Zuloaga09,Ciraci12,Wiener12,Stella13,Zhang14,Esteban15,Barbry15,Zapata15}
On the other hand, 
even if typical 
surface plasmon excitations localize 
in the nanometer scale, recent 
{\it ab initio} calculations considering realistic nanoparticle structures 
have shown that the fine atomistic details of the crystallographic facets and vertices of the metal particle, 
with the presence of single atomic protrusions and edges, 
introduce further non-resonant light localization.~\cite{Barbry15,Varas15}
This is analogue to the macroscopic
lightning rod effect,~\cite{Martin97,Li03} but brought down to the atomic scale.

It is thus timely to carry out a deep exploration 
on how the optical response of plasmonic cavities simultaneously correlates with 
their structural and transport properties,~\cite{Wen15} 
going beyond the macroscopic description to focus on the influence of 
strong atomic-scale structural 
reconfigurations of the cavity.~\cite{Varas15,Barbry15,Rossi15} 
This is important since, when two metallic surfaces are approached and put into contact, the formation
of small metal necks or nanojunctions connecting them is a very likely process,~\cite{Agrait03} 
and indeed this spontaneously happens
in our simulation. 
The formation of such metal nanocontacts has been theoretically predicted~\cite{Landman90,Agrait03}
and experimentally observed~\cite{Correia97,Kondo97,Kizuka98,Ohnishi98,Yanson98,Kondo00,Rodrigues00,Stoffler12,Agrait03}
These structures are at the root of friction phenomena in metal surfaces~\cite{Bhushan95,Landman96}
and give rise to quantized transport following discontinuous changes in the contact
cross-section.~\cite{Gimzewski87,Ferre88,Garcia89,Muller92,Todorov93,Agrait93,Pascual93,Olensen94,Torres96,Agrait03}
Thus, the key question that we want to address in this manuscript is whether a slight modification of
the geometry of the cavity, involving the movement of a few or even
a single atom in such sub-nanometric junction, e.g., due to migration or repositioning driven by
strain accumulation in a metal neck, 
and its corresponding change in conductance are clearly reflected in the optical response.  
If this were the case, 
one could expect to observe  a
discontinuous change in the plasmonic response of the system
accompanying each plastic deformation event. As shown below, our simulations
indeed confirm that such expectations are fulfilled. 

The present study is particularly relevant in the light of recent progress in fabrication and processing techniques. 
As the dimensions of nanoscale architectures are progressively reduced,
we are facing a regime where the actual distribution of the atoms 
in a system matters.~\cite{Ward07} 
The fact that Optics might follow the atoms 
is of utmost importance in optical engineering and optoelectronics, targeting optical modulators or electro-active control of optical signals, where instabilities and modifications of the performance can be attributed to atomic-scale features.~\cite{Emboras15} 

In order to address the complex correlation of electronics and optics in sub-nanometric junctions where the atoms 
in the system are allowed to adapt to the mechanical boundary conditions, 
we performed 
atomistic quantum mechanical calculations of the electronic structure,
the optical response, and the structural evolution of a plasmonic cavity. 
In our calculations we employed
an efficient implementation~\cite{Koval1,Koval2,Koval3} of linear response 
Time-Dependent Density Functional Theory (TDDFT) in conjunction with the 
SIESTA Density Functional Theory (DFT) package.~\cite{SIESTA1,SIESTA}
The plasmonic cavity in our simulation 
is formed by two large sodium clusters, containing 380 atoms each of them, 
in close proximity.  
This is a canonical example of a metallic 
system whose properties can be quite straightforwardly extrapolated, with care, to other metallic systems.  Further details of the calculations can be found in the Supplemental Material.~\cite{SuppMat}
The use of sodium allows performing larger calculations, in terms of the number of atoms, 
as compared to other more technologically 
relevant materials like, e.g., gold. This increases the relevance 
of our results, since the number of atoms involved in the structural reorganizations
of the neck is indeed a small 
percentage of those contained in the system. 

Our approach consists in tracing the energetics, the geometry and the optical response 
of the two metallic clusters which gradually get closer while allowing atoms to rearrange.
Remarkably, at a given separation distance, the clusters show a jump-to-contact instability~\cite{Pethica88}
leading to a strong modification of the optical response, a result 
in striking contrast to the smooth evolutions found in the context of previous 
classical and quantum descriptions based on  
static geometries.~\cite{Lassiter08,Zuloaga09,Esteban12,Zhang14,Barbry15}
The subsequent process of retraction of the clusters is particularly interesting. In such a situation, 
consistently with previous studies,~\cite{Sorensen98,Jelinek03} a metallic atom-sized contact 
is formed and the conductance across
the gap gets quantized. 
This allows revealing the strong correlation 
between the transport and optical properties of the system
and how, paradoxically, the motion of a few atoms or even a single atom in a nanometric gap can drive a 
quantized-like 
(abrupt, discontinuous) evolution of the optical response.

\section{Atomic rearrangements in the plasmonic junction: nanoparticles approach and retraction}

Our model of the plasmonic cavity is formed by 
 two sodium clusters that are progressively approached and retracted from each other.
Structural relaxations, using forces obtained from DFT calculations, 
are performed for each approaching and retraction step. Each cluster forming the dimer 
has an icosahedral shape and contains 380 sodium atoms.~\cite{SuppMat} 
The lateral dimension of each cluster is $\sim$25~$\ang$. 
The initial configuration consists of the two clusters placed at a distance of 16~$\ang$, 
a distance large enough to avoid direct interaction between the nanoparticles. 
The cavity is initially symmetric with the two clusters opposing planar facets.
Starting from that distance the two clusters are slowly brought together. 
We monitor the inter-particle distance using the {\it nominal gap size}, defined as 
the distance between the two cluster inner 
facets if the system would remain unrelaxed.
Thus, a nominal size gap of zero value 
would correspond to the superposition of the atoms forming the two 
opposing facets in the absence of relaxation. 
Approaching steps of 0.2~$\ang$ were chosen as a compromise between computational convenience and 
an approximately adiabatic 
evolution of the structure. 
As described in detail in the Supplemental Material,~\cite{SuppMat}
in order to control the clusters distance and to mimic the presence of tips or surfaces
the clusters are attached to,~\cite{Savage12} 
the atoms belonging to the outer facet of each cluster are kept fixed 
in their initial positions, i.e, they are not relaxed but move rigidly.
Once the distance between clusters corresponds to the typical inter-layer distance in bulk sodium, 
a process of retraction is started by pulling the clusters apart, again in steps of 0.2~$\ang$, until they completely separate.

\begin{figure}[h]
\centering
\includegraphics[width=8.5cm]{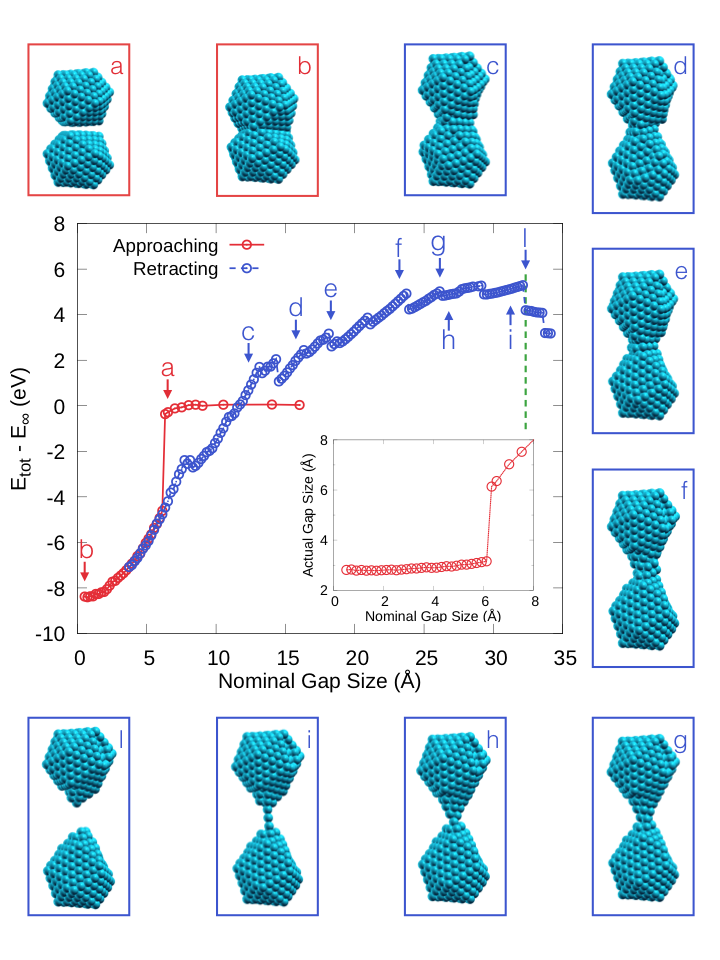}
\caption{Total energy of a plasmonic cavity formed by two Na$_{380}$ 
clusters as a function of the nominal distance between them ({\it nominal distance}
is defined as the distance among 
the facets defining the cavity if the system would not be relaxed). Red circles 
represent the approaching process while blue circles indicate the retracting process. 
The inner panel shows the actual gap size as a function of the nominal value during the 
approach, showing the clear signature of a jump-to-contact event. 
Images of the geometries around the graph show 
the rearrangement 
of the atoms in the clusters and the formation of a nanojunction during retraction. 
Latin letters indicate the correspondence between some of the total energy points in the graph and the  
configurations shown in the surrounding panels.}
\label{f:first}
\end{figure}

The total energy of the 
system during this process of approaching and retracting 
is shown as a function of the nominal gap size in Fig.~\ref{f:first}. 
The evolution of the clusters and junction geometry is shown in the panels of Fig.~\ref{f:first} for selected separation distances. 
The latin letters that label each panel relate the geometry of the system to the corresponding nominal gap size 
and energy, 
as indicated in the curves of the figure.
During the approach (red circles) the total energy remains constant until a nominal gap size of about 7.5~$\ang$ is reached. 
From this separation,  
the total energy starts decreasing smoothly. 
At a nominal gap size of $\sim$6.1~$\ang$ the two clusters suddenly jump to contact.~\cite{Pethica88}  
At this point the actual interface distance abruptly decreases and the energy is substantially reduced. 
The clusters are now connected and 
elongated along the intermolecular axis. 
The inset in Fig.~\ref{f:first} shows the actual gap size indicating the real distance between the 
inner facets of the clusters.
After the jump-to-contact, the value of the actual gap size drops to a value of about 3.2~$\ang$ remaining fairly constant and close to 3.0~$\ang$ as the two clusters get closer together.
The abrupt reduction of the energy at the jump-to-contact point is mostly due to the reduction
of the surface energy of both clusters (two facets disappear). However, this happens
at the expense of a large elastic deformation of the clusters.  
By further approaching the two clusters we reduce the elastic deformation of the system 
and, correspondingly, the total energy decreases. 
Eventually the system suffers some reorganizations which are also reflected (although they are somewhat
less obvious than the jump-to-contact) in the
energy versus distance curve in Fig.~\ref{f:first}. For example, 
the stacking of the atomic layers at the interface, 
initially imposed by the mirror symmetry of our starting geometry, gets optimized at a nominal 
gap size of approximately 2.5~$\ang$. Later, the particles start to deform to try to reduce further
their surface area by increasing the contact cross-section. The energy finally stabilizes and starts
to slowly increase for nominal gap sizes below $\sim$1~$\ang$. We stop our
approaching process at this point. 

Once the two clusters are clumped together at a nominal distance comparable 
to the interlayer distance in bulk sodium, 
we start pulling apart them (blue circles in Fig.~\ref{f:first}). 
During the retraction process the whole structure evolves creating and thinning a neck that connects the two clusters until a 
monatomic chain is formed and, eventually, until a complete separation of the clusters is achieved (point {\it l} in Fig. \ref{f:first}). 
In agreement with previous  studies, the evolution of the contact structure takes place 
via an alternation of elastic and plastic deformation events.~\cite{Rubio96,Torres96,Sanchez-Portal97,Untiedt97,Sorensen98,RubioBollinger01,Jelinek03,Agrait03}
The contact is elongated until the accumulated elastic energy is sufficient to 
produce atomic rearrangements, mainly driven by the atoms in the neck area. 
During these plastic events the
energy of the system decreases abruptly. Thus, there is a one-to-one correspondence
between the discontinuities of total energy in Fig.~\ref{f:first} and the changes in the configuration
of the metal neck. It is striking to note  
 the dramatic contrast between the distance at which 
the jump-to-contact takes place and
 the clusters ``touch'' for the first time during the approaching process (close to point a in Fig.~\ref{f:first}), 
 and the distance at which they finally detach  (indicated by a vertical green dashed line).
A nominal gap distance of 32.3~$\ang$ is needed to separate completely the clusters. 

In summary we have seen that the geometry of the system
strongly departs from the idealized situation in which 
two clusters simply change their relative position. These 
structural rearrangements had been overlooked
by most previous studies of plasmonics although, as we describe in the following, 
they play a key role in determining the optical response of the cavity.

\section{Optical response of a forming plasmonic cavity: relaxed vs. unrelaxed cases}

\begin{figure}[h!]
\centering
\includegraphics[width=7.cm]{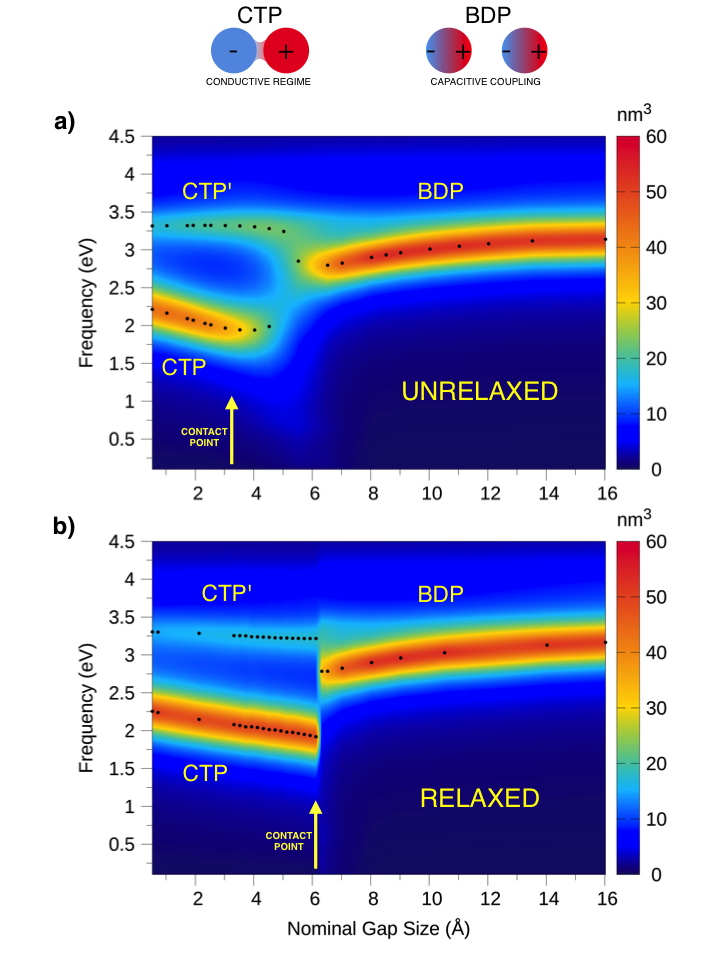}
\caption{
Evolution of the imaginary part of the polarizability of a Na$_{380}$ dimer 
(external field applied along the dimer axis)
as the clusters are approached, plotted as a function of the separation distance and 
photon energy. 
Both unrelaxed (a) and relaxed (b) geometries of the cluster dimer are considered.
The dark dots indicate the position of the peak maxima in the polarizability for those distances
for which the optical response has been computed.
The arrow lines indicate the {\it contact point} for the two cases, i.e., 
the distance at which the clusters merge into one single larger 
object. The top panel shows the schematic representation of the induced charge
in the modes that dominate the optical response before contact (bonding dipolar plasmon
mode, BDP) and after contact (charge transfer plasmon mode, CTP). Panels 
c) and d) show the imaginary part of the induced density and the corresponding modulus of the electron 
current flowing through each cross sectional [i.e., $(x,y)$] plane along the dimer axis (see the Supplemental 
Material for more details~\cite{SuppMat}). 
An external electric field of magnitude of $1\mathrm{x}10^{-9}$ atomic units is assumed here with a
polarization parallel to the junction main axis. The nominal gap size is 
6.1~$\ang$, corresponding to the jump-to-contact configuration in the relaxed case.
}
\label{f:second}
\end{figure}

To explore the connection between the structural evolution and the plasmonic response of the
cavity, we calculate the optical absorption of a plasmonic junction as the clusters approach for two situations: first, no relaxation of the clusters is allowed and the only parameter modified is the distance between the particles [Fig.~\ref{f:second}~(a)]; 
second, the relaxation of the atoms is taken into account 
[Fig.~\ref{f:second}~(b)] following the atomic-scale restructuring shown in 
Fig.~\ref{f:first} (red symbols).
The resonant plasmonic modes of the forming cavity, as obtained from the calculated  polarizability of the system,  
are displayed as a function of the inter-particle distance in both situations.
The component of the polarizability parallel to the dimer axis is considered here, i.e., induced by an electrical field polarized along the same axis, 
that we take hereafter as z. 
In both cases, and depending on the separation, we can 
identify three distinct resonances.~\cite{PerezGonzalez10,Zuloaga09,Esteban12,Savage12,Zhang14,Esteban15,Barbry15}
 A single intense so-called Bonding Dipolar Plasmonic (BDP) resonance around 3 eV dominates the response 
at large inter-cluster distances when the two clusters interact weakly.
The BDP shows an induced charge distribution characterized by a capacitive coupling of charges of opposite sign at both sides of the cavity, as schematically depicted in the 
right drawing of the top panel in Fig.~{\ref{f:second}. 
When both clusters are in contact, so that free charges can efficiently move across the junction, 
we enter a conductive coupling regime characterized by the so-called Charge Transfer Plasmon (CTP) and the associated high-energy Charge Transfer Plasmon (CTP') modes. 
The conducting link of the CTP through the junction of the clusters, produces a screening of the charges in the cavity, and thus 
redistributes the induced charge density to produce a net dipole that extends 
to the whole dimer structure, as depicted in the 
top-left scheme in Fig. \ref{f:second} [see also the panel c) and d) ].

The BDP resonance is redshifted as the inter-cluster distance is reduced and the Coulomb interactions among the clusters increase. 
This shift is due to the strong
interaction of the parallel induced dipoles along the dimer axis, which hybridize~\cite{Prodan03} 
lowering the energy of the resulting optically active mode. 
In this capacitive (weak interaction) 
regime both unrelaxed and relaxed cases show the same dependence on the
inter-particle distance.
The BDP mode lives until the clusters are brought to a distance of about 6.1~$\ang$. 
At this point, for the unrelaxed dimer [see Fig.~\ref{f:second}~(a)], 
the BDP mode is quenched and higher energy modes
start gaining intensity. 
If the clusters are approached further we observe a smooth transition from the
capacitive to the conductive coupling regime. 
For separation distances right below 6~$\ang$ the electron tunnelling current at relevant frequencies 
gradually starts flowing, giving rise to the progressive emergence of the 
CTP resonance.~\cite{Marinica12,Esteban12,Esteban15, Barbry15} 
This transition region is frequently referred to as the quantum tunnelling regime of
plasmonic cavities.~\cite{Savage12}
At a distance comparable to the sodium interlayer distance, $\sim$3.0~$\ang$, the clusters 
become chemically bonded 
and a clear contact is established. Under those conditions, a substantial current can be established and the CTP 
appears fully developed.  

In contrast, the situation shown in Fig.~\ref{f:second}(b) for the relaxed dimer is strikingly different.
The relaxed dimer 
undergoes a jump-to-contact instability (see red curve in Fig.~\ref{f:first})
that dramatically modifies the evolution of the optical spectrum. 
The transition regime, found between 6~$\ang$ and 3~$\ang$ for the unrelaxed dimer, has almost completely 
disappeared in the relaxed case. There are not stable geometries for those intermediate
gap sizes and, thus, the resistive tunnelling (transition) regime cannot be clearly identified
in the optical response in this case.
Although the details of the jump-to-contact process strongly depend on the size and shape of the facets and the 
effective elastic constants of the systems being brought into contact, this effect is a quite general behavior which is routinely 
taken into account in the interpretation of data from scanning probe microscopies.~\cite{Hofer03}
Our results indicate that the effect of the jump-to-contact must be considered when exploring and interpreting the optical response 
of metallic particles in close proximity, particularly when large atomic-scale reconfigurations 
can be expected. Importantly, this phenomenon can hinder the appearance of a smooth transition between the capacitive and charge-transfer regimes 
in the optical response of plasmonic cavities. 

In panels (c) and (d) of Fig.~\ref{f:second} we explore the real space distribution of the induced charge 
for the CTP and CTP' modes right after the clusters get into contact, i.e., 
right after the jump-to-contact instability.
Here we plot the imaginary part of the induced density at the resonant frequencies.
We also plot the corresponding electron current (graph to the right of 
each charge density plot) flowing through $(x,y)$ planes (i.e., 
perpendicular to the dimer axis) as function of $z$, the coordinate along the dimer axis.
Details of the method used to calculate the current can be found in the 
Supplemental Material.~\cite{SuppMat}
The density charge associated with the CTP forms a dipolar pattern over the whole system having a single node placed at the center of the system. Thus, the charge accumulation does not take place
in the cavity interfaces, but rather extends to the whole system. Correspondingly, the  
current
associated with the CTP resonance has its maximum at the gap center. 
On the other hand, the CTP' mode presents two dipolar patterns on each cluster with nodes of the 
induced density charge in the center of the system as well as in the middle of each cluster. The 
charge distribution in this case is somewhat similar to what one can expect for the 
BDP mode. However, 
the current reveals a key piece of information to rule out this interpretation. 
In the case of the CTP' resonance the maxima of the current are found both
in the center of the system as well as within each cluster. This is indeed confirming
that there is charge transfer among both clusters also in this high energy mode.
Thus, the observed
induced density pattern is better interpreted as  the second optically active mode of 
a metal rod. 

Finally, below 2~$\ang$ nominal gap size 
the conductive coupling regime 
of the junction is fully developed in both the unrelaxed and relaxed cases, with 
the CTP and the CTP' resonances converging to similar values of energy, around 2.25 eV and 3.3 eV, respectively.
This underlines the fact that the details of atom rearrangements at the cluster interface 
might not be so important in the determination of the optical response once the two clusters are fully chemically bonded.

\section{Optical response of a retracting plasmonic junction: Optics driven by individual atoms}

\begin{figure*}[htpb]
\centering
\includegraphics[width=15cm]{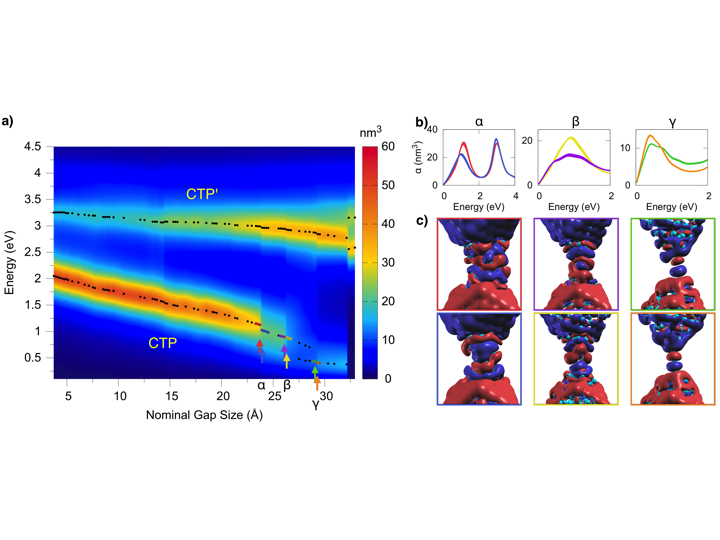}
\caption{a) Evolution of the resonances in the polarizability of a plasmonic junction 
as a function of the nominal gap size and energy, 
as the clusters forming the junction move apart (i.e., move towards larger nominal gap sizes). 
The dots indicate the positions of the peak maxima in the polarizability for the considered configurations.
Panels in b) show the spectral lines of the polarizability at distances before and after 
each of the jumps highlighted in panel 
a) by means of Greek letters and colored arrows and dots. 
The colors of the lines of the spectra correspond 
to the colors of those arrows and dots, thus indicating whether a given spectrum corresponds to a configuration before or after the jump. 
c) Imaginary part of the induced charge density distribution around the junction 
for the three selected distances in a) and b), before (top) and after (bottom) the jumps. 
}
\label{f:third}
\end{figure*}

Fig.~\ref{f:third} shows the optical polarizability of the junction during the retraction process.
Surprisingly, as the two clusters are retracted, the CTP and CTP' modes 
dominate the spectrum for most separation distances, all the way to nominal inter-particle distances of several tens of Angstroms. 
This is in clear contrast with the results obtained in the previous section (approaching situation), and 
it is a result
of the structural evolution of the junction, characterized by 
the formation of a thin conducting neck among the clusters
[as shown in the panels (c)-(i) of Fig.~\ref{f:first}].
As the clusters get separated, the neck gets longer and thinner.
As a consequence, 
the charge transfer modes disperse towards lower energies (due to the overall elongation 
of the system). 
Moreover, as the neck cross-section is reduced, the intensity of the CTP' mode increases 
at the expense of the lower-energy CTP resonance, consistent with calculations of 
stretched clusters.~\cite{Rossi15} 
As the current flowing across the neck diminishes, the CTP' mode converges towards
the BDP mode while the CTP mode tends to disappear.

While the polarizability of the 
approaching situation in Fig.~\ref{f:second} only shows a clear discontinuity 
associated with the jump-to-contact instability of the cavity, the 
retracting situation shows a completely 
different behavior as a function of the separation distance. 
During retraction the 
optical spectrum is characterized by the appearance of many discontinuities both 
in the spectral position and the intensity of the resonances. 
A careful inspection of Fig.~\ref{f:first} reveals that 
these discontinuities happen at exactly the same nominal distances 
where jumps in the total energy are detected.
Some of the most visible jumps are highlighted with arrows of different colors and 
marked with Greek letters in Fig.~\ref{f:third}~(a), and the corresponding polarizability 
is plotted in detail on the panels of Fig.~\ref{f:third}~(b), which extracts the spectral 
lines from the contour plot in (a). 
Each panel shows spectra corresponding to distances before and after one of the jumps, 
identified in the contour plot of Fig.~\ref{f:third}~(a) with the corresponding colored dots and
arrows. 
Consecutive curves correspond to configurations in which the inter-particle distance is
changed by 0.2~$\ang$. In each panel there are several, almost indistinguishable, 
spectra of the same color. This highlights that noticeable changes in the spectrum
are indeed linked to the plastic deformation
events in the neck, and not to the small rearrangements during elastic deformation. 
At each jump we observe clear changes in the intensities, widths
and positions of the resonance peaks. 
The jumps affect primarily the low energy resonance, CTP, although they are also visible in the
CTP' mode. 
They are owing to the atomic reorganization in the neck region and they are specially 
visible for distances above $\sim$20~$\ang$ 
due to the small cross-section of the neck. 
Remarkably, for such thin necks even single atom movements produce visible changes in the optical response of the system, clearly associated with the quantized nature of the conductance through the junction neck. 
The jump at 29.3 $\ang$ indicates the formation of a well-ordered
monatomic neck, i.e., the clusters are connected by a single row of atoms. 
The formation of such structures has been observed for many metals, for example in the case of Au, 
for which these monatomic wires have even 
been visualized by electron 
microscopy~\cite{Ohnishi98,Yanson98,Sorensen98,Sanchez-Portal99,RubioBollinger01,Rodrigues01,Koizumi01,Kizuka08,Agrait03}.

The intensity of the CTP resonance suffers an abrupt decrease for 
a nominal gap size of around 23~$\ang$, becoming broader between 27 and 29.3~$\ang$.
Afterwards, simultaneously to the monatomic neck formation, the CTP resonance gets sharper with a consequent regain in intensity.
This evolution is due to a combination of several effects, the most important being the 
quantization of electron transport in the metal neck. Such quantization 
is a well-known effect due 
to the small cross-section of the contact, comparable to the electron 
wavelength.~\cite{Dattabook} 
As a result of the lateral confinement, 
the electronic energy levels in a thin metal nanowire or neck get quantized and, at a given energy, only a discrete number of bands (or ``channels'' using the standard terminology in quantum transport) 
can contribute to the electron transport. Thus, under a small, static bias, 
if the electron injection from the electrodes (in our case the clusters) is efficient and the neck structure is sufficiently long and ordered, we
can expect each channel at the Fermi level 
to contribute to transport with a quantum of conductance $\mathrm{G}_0=2e^2/h$,~\cite{Dattabook}
with $h$ the Planck's constant. In the presence of defects or strong scattering in the connections to 
the electrodes, the transmission probability of the channels gets reduced.~\cite{Dattabook}

With these ideas at hand, we can easily explain the observed behaviors. 
The abrupt jump in the intensity of the CTP peak around 23~$\ang$ ($\alpha$ jump) is caused
by the sudden reduction of the neck's cross-section, as can be clearly seen in Fig.~\ref{f:third}~(c) and 
the inset of Fig.~\ref{f:fourth}. 
As expected, the reduction of the cross-section reduces the number of
conduction channels and, therefore, the electric current flowing through the junction (this is confirmed
in Fig.~\ref{f:fourth}, discussed later in detail). The resonance peak also shifts to slightly lower energies. 
The origin of the intensity jump at $\sim$27~$\ang$ ($\beta$ jump) is also similar:
a cross-section reduction that translates onto a sudden decrease of the current as can be 
seen in Figures \ref{f:third}~(c) and \ref{f:fourth}. 
After this jump at 27~$\ang$, the neck develops into a less ordered structure,  
creating a region of high scattering that hampers the electron 
transport between the clusters. As a consequence the CTP resonance broadens. 
Finally, once the relatively defect-free monatomic wire is formed, the transport through the neck becomes completely ballistic, 
i.e. all the electron that are injected to the monatomic wire get across the junction, and the peak in the polarizability becomes more defined again.  

These quantization effects can  also be observed in 
the shape of the 
distributions of induced charge density as the neck 
evolves during retraction. 
In panels (c) of Fig. \ref{f:third} the imaginary part of the induced density associated with the CTP mode 
is plotted for those configurations immediately 
before and after the $\alpha$, $\beta$ and $\gamma$ jumps 
(indicated
by the colored arrows in the polarizability plot). 
Although the density change has a quite complex distribution, it is possible to follow the evolution of the patterns 
towards simpler schemes of charge oscillation after each jump. 
The induced density presents a complex distribution and nodal structure, with a decreasing number of nodes as the cross-section 
of the neck gets thinner, a fact that reflects the larger number of open conduction channels for the wider structures. 
Subtle changes in the structure that have a direct translation in the optical response 
can also  
be observed in these density plots. 
For example, in the case of the $\gamma$ jump,
the three-atoms-long monatomic wire becomes more straight 
and the connections to the cluster more symmetric. 
This slightly increases the current flowing through the structure and produces the aforementioned changes in the plasmonic response.  

Once the two clusters totally separate, breaking the neck, two split resonances arise near the CTP' resonance at about 2.6 and 3.2~eV. 
The initial face-to-face configuration has been substituted by an asymmetric 
tip-to-tip configuration (see panel {\it l} in Figure \ref{f:first}). 
The lower energy resonance recalls a BDP mode, with 
the largest charge accumulations around the central gap
(see  the Supplemental Material~\cite{SuppMat}).
The 
higher energy resonance has a more complex charge distribution, 
corresponding to higher order mode, showing
charge accumulations 
both in the tips inside the cavity and in the facets of the clusters
(see the Supplemental Material~\cite{SuppMat}).

\begin{figure}[htpb]
\centering
\includegraphics[width=8cm]{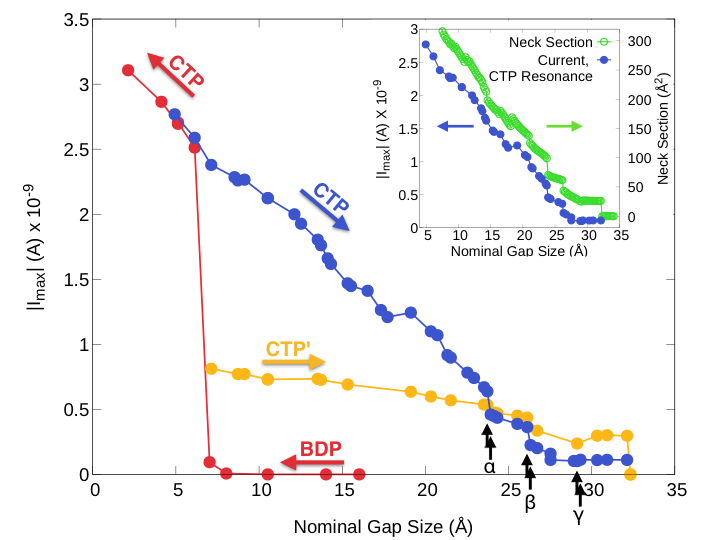}
\caption{Modulus of the current flowing 
through a plasmonic junction as a function of the separation of the clusters forming the junction. The current is evaluated at 
a cross-sectional plane
passing through the center of the cavity (and cutting the center of the neck when present).  
Colored arrows indicate the direction of the process (approaching or retracting).
The current is computed at the resonance frequency of the different modes of the cavity, 
as indicated by the labels and described in the text.   
Black arrows indicate the position at which the spectral jumps in Figure~\ref{f:third} occur.
The inset shows the one-to-one correspondence between the jumps in the current for 
the CTP mode and the cross-section of the metal neck. An external electric field 
of 1$\mathrm{x}10^{-9}$ atomic units is assumed with a
polarization parallel to the junction main axis.
}
\label{f:fourth}
\end{figure}

To fully account for the connection between 
high-frequency electron transport and optical response of the plasmonic junction, 
we have calculated the current through the junction as a function 
of the nominal gap size. 
In Figure~\ref{f:fourth} the results for the modulus 
of the current passing through a plane cutting the center 
of the junction are shown. 
Here we present the current computed at the frequencies of the main resonances of the polarizability, described in previous sections. 

The current during the approach process 
is shown by red solid circles, corresponding first to the BDP mode, and later 
to the CTP mode, once the clusters are in contact.
The current for the BDP mode is negligible until the jump-to-contact 
event takes place.
Once the clusters are connected the current can flow through the whole system and therefore its value increases dramatically. 
The current calculated for the CTP resonance increases almost linearly as we decrease the nominal gap size. 

The values of the current across the junction at the energies of the CTP and CTP' resonances, while retracting the clusters and the neck is getting thinner, are plotted respectively in blue and yellow.
The current related to the CTP resonance decreases monotonously as we  
elongate the system. As commented above, its evolution is characterized by abrupt jumps
whenever the neck suffers a plastic deformation. 
The current eventually reaches a plateau associated with the formation of a well-defined 
monatomic neck. 
Interestingly, once the monatomic neck is formed, a further neck stretching does not affect considerably the current. 
This can be expected since the conductance of such small necks mostly depends on the cross-section,
which is fixed for the monatomic wire. 
It is interesting to mention that for the low energy CTP mode, it was possible to define the conductance of the monatomic neck
as the ratio of the current and the voltage drop 
across the wire.
The modulus of the optical conductance (which is a complex number now) computed 
in this way at the CTP frequency is $\sim$0.65~$\mathrm {G}_0$, 
close to the expected value for a monatomic sodium wire at low frequency, $\mathrm {G}_0$. 
In this case 
it was possible to define the conductance without ambiguity 
since the potential drop is confined to the gap region, 
with the electrostatic potential flat inside the clusters.~\cite{SuppMat}
For larger cross-sections of the necks and/or higher energies of the mode, this condition is not fulfilled and the definition 
of the conductance is quite involved. 
The current associated with the CTP'  mode follows the same trends than that
 of the CTP, although shows a 
less pronounced dependence on the overall elongation 
of the system.
Obviously, once the clusters separate the current becomes negligible.

The arrows in Figure \ref{f:fourth} indicate the position of the jumps shown in Figure \ref{f:third}~(a). 
Except for the last jump at 29.3~$\ang$, 
the other two jumps observed in the polarizability (Fig.~\ref{f:third}) and in the total energy (Fig.~\ref{f:first})  
show a clearly correlated sudden change in the current. 
This points towards a remarkable effect of a few atoms (or even a single atom), whose motion
influences the overall optical response of the whole system (containing 760 atoms in our case). 
This observation can be of utmost importance in the control and manipulation 
of optical signal in subnanometric junctions which are clearly affected by this type of physical processes at the atomic scale. 

To establish a more direct connection between the computed current as a function of
the gap separation within the junction and the well-known quantization of transport in metal nanocontacts,  
in the inset panel of Fig.~\ref{f:fourth} we show the current for the CTP while retracting (left axis of the graph)  plotted in relationship with the neck cross-section (right axis of the graph). As can be seen,
there is an almost perfect correlation between the changes in the current and the evolution 
of the neck cross-section (see the Supplemental Material for the procedure adopted here 
to calculate the neck cross-section~\cite{SuppMat}). 
Such correlation has been already well-established in the case of low-frequency 
driving-fields being applied to the necks. It has been 
observed in the formation of metal nanocontacts in Scanning Tunnelling Microscopy (STM) 
and break-junctions experiments, and corroborated by many calculations.~\cite{Agrait03}
Our Fig.~\ref{f:fourth} goes one step beyond, establishing such correlation at optical 
frequencies. With this additional piece of information, we can now summarize the
results described in this section by unambiguously establishing the following cause-effect relationships:
plastic deformation of the neck during elongation 
$\rightarrow$ cross-section reduction $\rightarrow$ abrupt drop of
the current $\rightarrow$  decrease of the intensity of the CTP mode. This process
is sometimes accompanied by small shifts of the position of the resonance peak. 
Finally, as mentioned above, more disordered structures translate into
broader and dimmer CTP resonances.

\section{Conclusions}

In summary, we have shown how 
atomic-scale structural reorganizations are crucial to determine the optical properties of 
plasmonic cavities. Besides the importance of jump-to-contact events, that can
almost completely eliminate any signature of the plasmonic tunnelling regime, the  
effects are particularly dramatic 
when a metal nanocontact is formed across the cavity. This is due to 
the strong dependence of the plasmonic response of the system 
on the quantized current 
flowing through the connecting neck.

The mechanical response of atom-sized necks is characterized by sudden rearrangements of the atomic structure,
which frequently involve just a few atoms in the thinner part of the contact.
Since the electron transport through  thin metal nanocontacts is quantized, the corresponding changes of the current 
flowing across the junction are necessarily discontinuous. 
Our calculations demonstrate that this common observation under small  applied dc biases can  
be extrapolated to the optical frequencies of the plasmon resonances of the cavity, at least 
for the short ballistic contacts considered here. 
These jumps in the current translate onto abrupt changes in the 
plasmonic response of the system.  
Thus, the discontinuous evolution of the spectral position, width and intensity of the CTP mode observed
in our simulations is a direct consequence of the transport quantization in the connecting neck. 

The correlation is clearly demonstrated, showing that 
remarkably, Optics follows the atoms. 
This is absolutely important in the design of subnanometric-scale optical modulators that rely in slight changes of 
the optical response against tiny configurational modifications. 
In our case we have 
analyzed relatively small icosahedral sodium clusters, however, we expect to find a similar behavior 
for other materials suitable for electronic applications, such as gold. 

The effect of a single atom in the optical properties of a nanoscopic object as the one reported here, which 
can be probably extended to somewhat larger objects, has important consequences in optical engineering, molecular electronics, and photochemistry, where 
the optical response can now be tailored by a few atoms.

\begin{acknowledgements}
We acknowledge useful discussions 
with Andrei G. Borisov and Ruben Esteban regarding the role of electron transport in the determination 
of the optical properties of metal nanojunctions, and Dietrich Foerster regarding efficient TDDFT calculations.
We acknowledge financial support from projects FIS2013-14481-P and MAT2013-46593-C6-2-P from MINECO. 
MB, PK, FM and DSP also acknowledge support from the ANR-ORGAVOLT project and the
Euror\'egion Aquitaine-Euskadi program. MB acknowledges support from the Departamento 
de Educaci\'on of the Basque Government through a PhD grant. PK acknowledges financial support from 
the Fellows Gipuzkoa program of the Gipuzkoako Foru Aldundia through the FEDER funding scheme of the European Union. 
\end{acknowledgements}

\bibliography{Marchesin_etal}

\end{document}



\title{Supplemental Material for ``Optical response of metallic nanojunctions driven by single atom motion''}

\author{F. Marchesin}
\affiliation{Centro de F\'isica de Materiales CSIC-UPV/EHU, Paseo Manuel de Lardizabal 4, 20018 Donostia-San Sebasti\'an, Spain}
\affiliation{Donostia International Physics Center (DIPC), Paseo Manuel de Lardizabal 5, 2001 Donostia-San Sebasti\'an, Spain}
\author{P.Koval}
\affiliation{Donostia International Physics Center (DIPC), Paseo Manuel de Lardizabal 5, 2001 Donostia-San Sebasti\'an, Spain}
\author{M. Barbry}
\affiliation{Centro de F\'isica de Materiales CSIC-UPV/EHU, Paseo Manuel de Lardizabal 4, 20018 Donostia-San Sebasti\'an, Spain}
\affiliation{Donostia International Physics Center (DIPC), Paseo Manuel de Lardizabal 5, 2001 Donostia-San Sebasti\'an, Spain}
\author{J. Aizpurua}
\email{aizpurua@ehu.eus} 
\affiliation{Centro de F\'isica de Materiales CSIC-UPV/EHU, Paseo Manuel de Lardizabal 4, 20018 Donostia-San Sebasti\'an, Spain}
\affiliation{Donostia International Physics Center (DIPC), Paseo Manuel de Lardizabal 5, 2001 Donostia-San Sebasti\'an, Spain}
\author{D. S\'anchez-Portal}
\email{sqbsapod@ehu.eus}
\affiliation{Centro de F\'isica de Materiales CSIC-UPV/EHU, Paseo Manuel de Lardizabal 4, 20018 Donostia-San Sebasti\'an, Spain}
\affiliation{Donostia International Physics Center (DIPC), Paseo Manuel de Lardizabal 5, 2001 Donostia-San Sebasti\'an, Spain}


\maketitle

\section*{Methods: Structural Relaxations}

The structural relaxations of the Na$_{380}$
dimers were performed using standard density functional theory (DFT) 
as implemented in the SIESTA code.~\cite{SIESTA1,SIESTA}. 
We used the Perdew-Burke-Erzenhorf density functional (GGA-PBE)~\cite{Perdew96},  
norm-conserving pseudopotentials~\cite{TMpseudos} to effectively
describe core electrons, and a 
double-$\zeta$ polarized (DZP) basis set of numerical atomic orbitals generated using an
{\it energy shift} of 100~meV.~\cite{Artacho99} The fineness of the real-space grid used
to compute the Hartree and exchange-correlation contributions to the energy and Hamiltonian
corresponds to a plane-wave cut-off of 130~Ry.~\cite{SIESTA}
Structural relaxations were stopped when forces acting on all atoms were smaller than 0.02 eV/$\ang$. 

\begin{figure}[htpb]
\centering
\includegraphics[width=5cm]{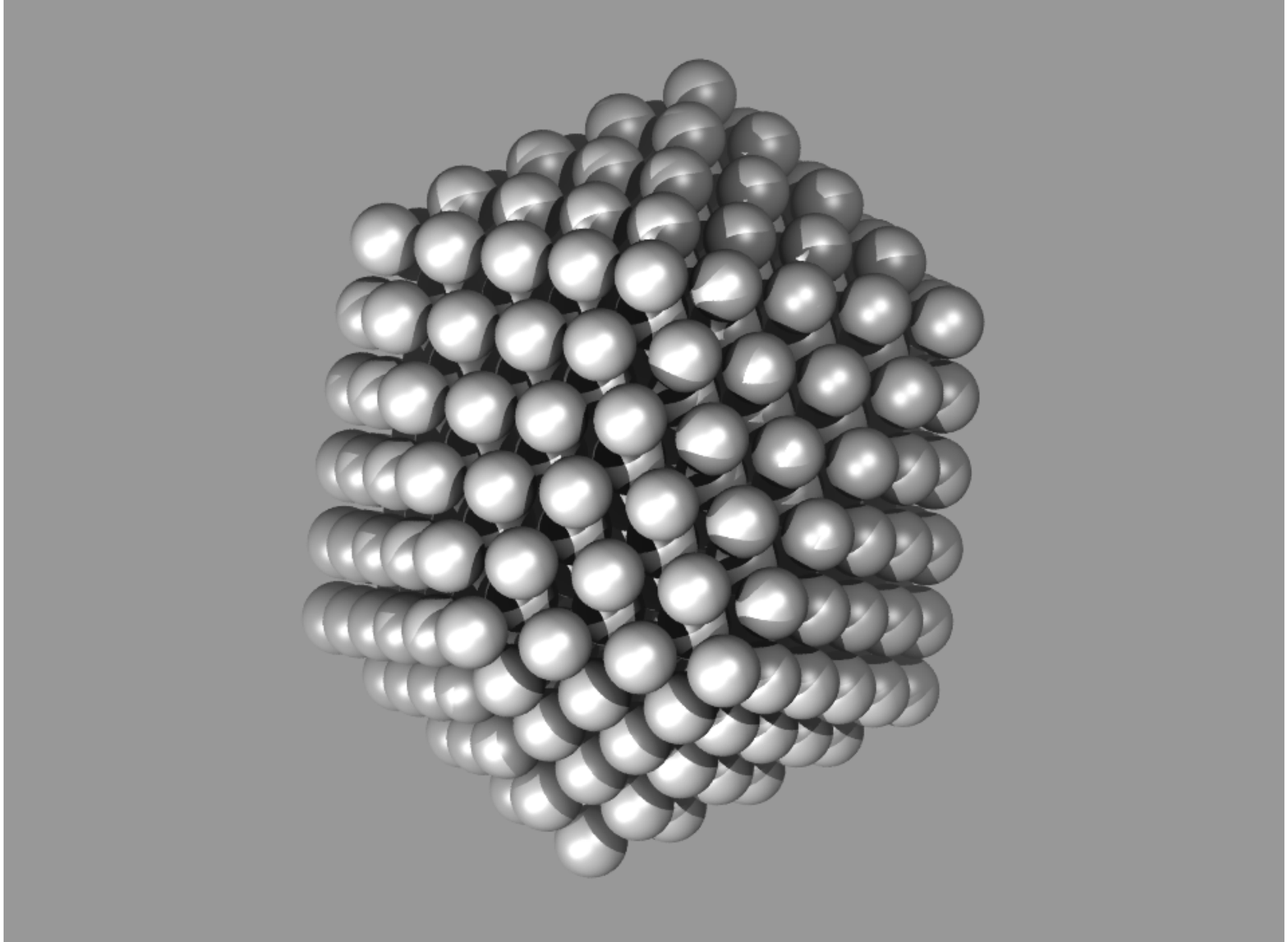}
\caption{Relaxed initial structure of one of the Na$_{380}$ clusters 
forming the plasmonic cavity.}
\label{FigureS1}
\end{figure}

Figure~\ref{FigureS1} 
shows the relaxed initial structure of the Na$_{380}$ clusters
that form the plasmonic cavity. 
The cluster presents an icosahedral structure and has been obtained
starting from a configuration optimized with empirical potentials~\cite{MM90}
and available at the Cambridge Cluster Database (CCD).~\cite{CCD}
The relaxed structure  obtained using GGA-PBE is very similar to the
one provided at the CCD site, and it is characterized by the presence of planar
facets, sharp edges and single-atom vertices. 

\begin{figure}[htpb]
\centering
\includegraphics[width=5cm]{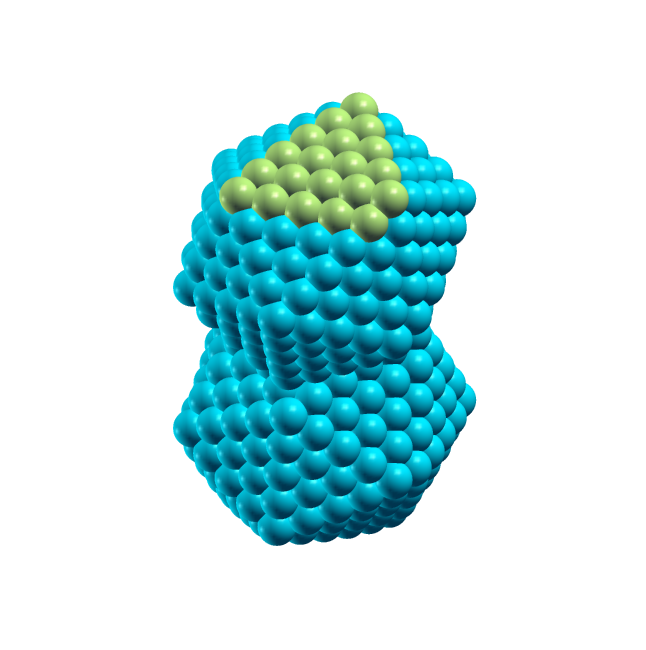}
\caption{The atoms in the outer facets of both clusters (the atoms in one of these facets have been highlighted
in yellow) are kept fixed during the relaxation process.}
\label{FigureS2}
\end{figure}

Figure~\ref{FigureS2} shows a snapshot of the relaxation of the plasmonic cavity 
when the to clusters are approached towards each other. As a mechanical constraint, 
and mimicking the presence of macroscopic scanning probe tips or surfaces where the
clusters are attached, the atoms in the outer surfaces of both clusters (highlighted in yellow
in Fig.~\ref{FigureS2}) 
remain fixed and move rigidly during the approach/retraction events. 
Each new configuration is generated by changing 
the distance between these fixed facets by 0.2~$\ang$.
The rest of the atoms, blue atoms in Fig.~\ref{FigureS2}, are then relaxed. 
Steps of 0.2~$\ang$ are chosen as a compromise between numerical efficiency and 
an approximately adiabatic evolution of the system. The presence of two unrelaxed facets
allows to measure the displacement applied to the system unambiguously, which in turn 
permits the definition of a {\it nominal gap size}, corresponding to the distance between 
the inner facets of the neighboring clusters in case the system were not allowed to relax. 
This is the magnitude that is used to define the inter-particle distance in the 
plots in the main text of our paper. 

\section*{Methods: Optical Response Calculations}

The calculations of the optical response of the plasmonic cavity were performed within the framework of 
time-dependent density functional theory (TDDFT) using the so-called adiabatic local-density approximation (A-TDLDA), 
i.e., using the LDA~\cite{LDA,LDA2} exchange-correlation kernel computed for the instantaneous electron density.
TDDFT is a rigorous (in principle) extension of DFT to time dependent systems.~\cite{runge84}
We use an 
efficient implementation of linear response TDDFT that utilizes an iterative scheme to compute the optical response in frequency domain and a local basis to expand the products of Kohn-Sham orbitals.~\cite{Koval1,Koval2,Koval3} The program uses as an input the information from ground-state electronic-structure calculations performed using localized basis sets (e.g.,  linear combination of atomic orbitals). In particular,  the method has been interfaced with SIESTA and, as exemplified here, allows computing the linear optical response of large clusters and molecules with moderate computational resources.~\cite{ACSNano13,Barbry15}.
We imposed an intrinsic (phenomenological) broadening $\Delta=300$~meV, 
which brings the total width of the calculated 
plasmonic resonances for the single cluster in reasonable agreement with available 
experimental data.~\cite{Schmidt99} The frequency axis is sampled with a resolution of 0.02~eV.

Before computing the polarizability, for each relaxed structure, a DFT ground-state calculation has to be 
performed employing the LDA functional. The resulting Kohn-Sham orbitals and energies
are used as an input for the subsequent A-TDLDA calculations.
While the use of the GGA functional substantially improves the description of the geometries for sodium clusters 
(in particular, LDA tends to produce too dense structures,~\cite{Cohen_Na} 
which in turns can give rise to a blueshift of plasmon
resonances~\cite{Barbry15}), A-TDLDA has proven to provide accurate results for the optical properties
of sodium clusters.~\cite{ARubio96,Vasiliev99,Vasiliev02,Tsolakidis,Marques06,Alonso06}

\section*{Methods: Calculation of Electric Currents at Optical Frequencies}

\begin{figure}[htpb]
\centering
\includegraphics[width=4cm]{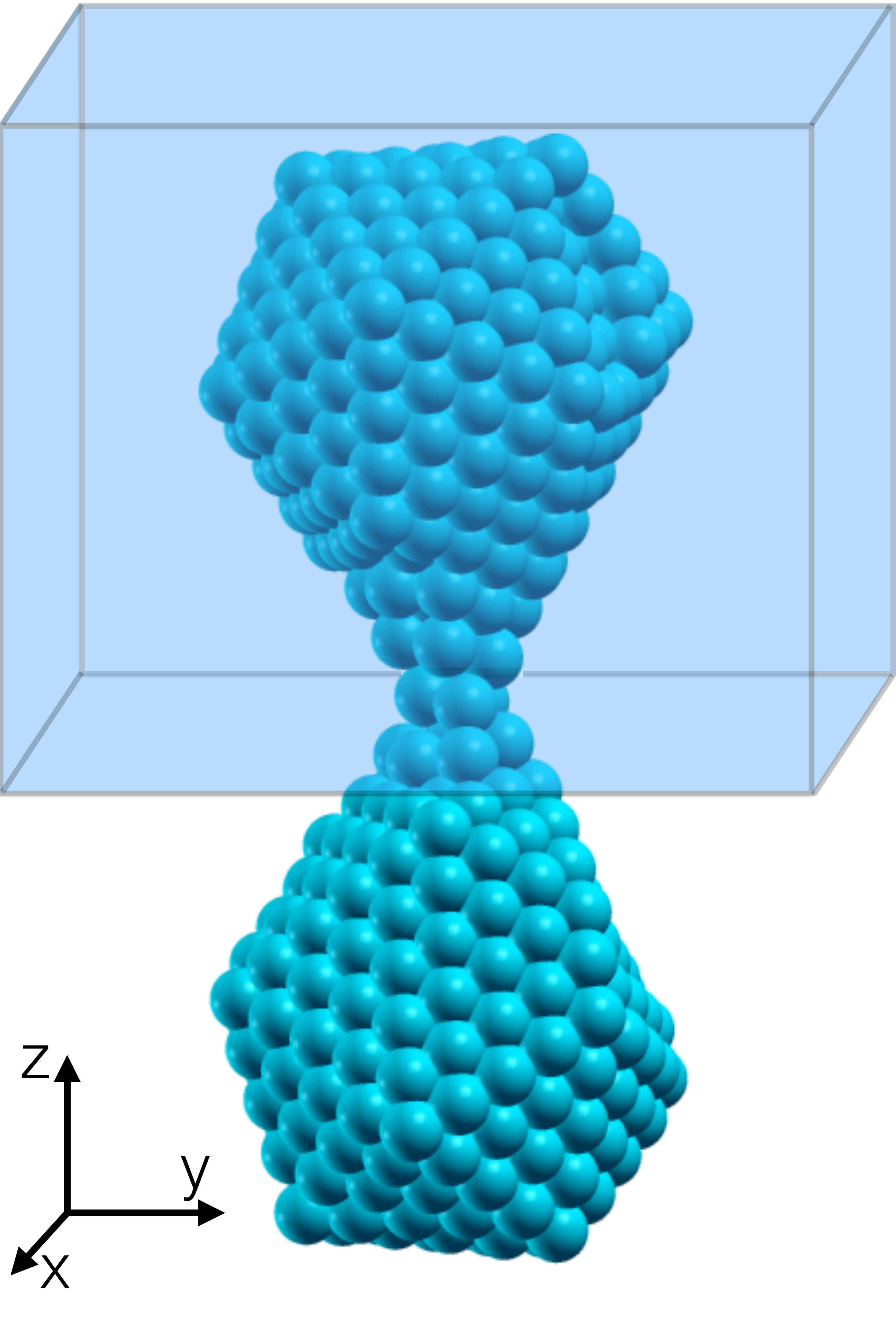}
\caption{Representation of the integrating volume used to define the current passing through 
the junction with the help of the continuity equation. }
\label{FigureS3}
\end{figure}

Since we are dealing with a finite object we can use the continuity equation and an integration region like 
the one shown in Fig.~\ref{FigureS3} to define the current that flows across a plane perpendicular to
dimer axis and passing through the center of the junction. From our TDDFT calculation we obtain the
induced electron density $\delta \tilde{n}(\bf{r},\omega)$ in response to a monochromatic field with frequency $\omega$.
We can now integrate the induced density over the volume in Fig.~\ref{FigureS3} 
 \begin{equation}
 \delta\tilde{Q}(\omega)=\int_{\Omega}  dr^3  \delta \tilde{n}(\bf{r},\omega),
\label{e:totalcharge}
\end{equation}
to obtain $\delta\tilde{Q}(\omega)$ the Fourier transform of the time-dependent total induced charge contained in $\Omega$. 
The continuity equation provides with the trivial relation between the total induced charge in $\Omega$, $\delta Q(t)$, and the current
flowing across the junction $I(t)$ at each instant of time $t$. In frequency 
domain this relation is expressed as
 \begin{equation}
 \tilde{I}(\omega)= -i\omega \delta\tilde{Q}(\omega).
\label{e:continuity}
\end{equation}
Thus, we can easily obtain the modulus of the current (maximal current) flowing through the junction in response to
an external electric field of the form ${\bf E} = {\bf E}_0 cos(\omega t)$:
\begin{equation}
|I_{max}(\omega)| = \omega\sqrt{\left[ \delta \tilde{Q}'(\omega)\right]^2 + \left[\delta \tilde{Q}''(\omega)\right]^2 },
\label{e:I_max}
\end{equation}
with  $\delta \tilde{Q}'(\omega)$ and $\delta \tilde{Q}''(\omega)$, respectively, the real and imaginary 
parts of $\delta\tilde{Q}(\omega)$.

In our calculations we compute the current at the frequencies of the plasmonic resonances for
each geometry,  and 
$\delta \tilde{Q}(\omega)$ is calculate by integrating the density change $\delta \tilde{n}({\bf r},\omega)$
over a given real space volume ($\Omega$). 
By changing the integration volume we define the $(x,y)$ plane through which the current is 
calculated. For example, the
current as a function of the position of such plane is shown, for a particular configuration 
and two different resonance frequencies, in the panels c) and d) in Figure~2 of our paper.
It is important to stress that such maps do not refer to a given instant of time, but 
they rather depict the maximum current passing through each $(x,y)$ plane at any instant of time.
In order to compute the current flowing through the center 
of the junction (as shown of Fig.4 of the main text) 
we need to integrate the induced charge over a volume similar to that depicted in Figure \ref{FigureS3}.

\section*{Methods: Definition of the Neck Cross-Section}

\begin{figure}[htpb] 
\centering
\includegraphics[width=5cm]{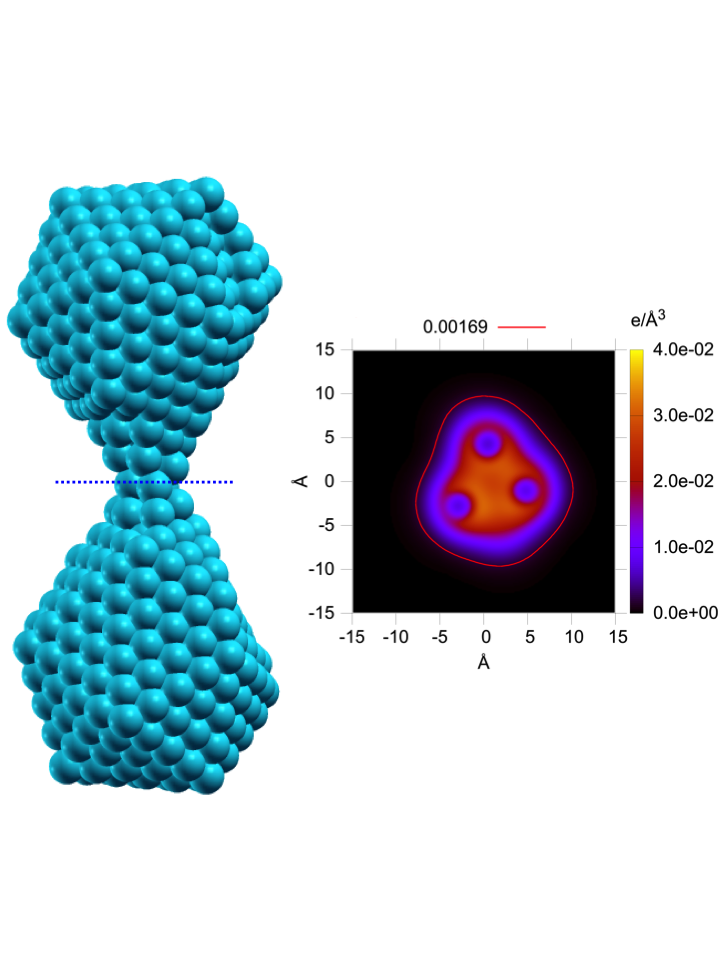}
\caption{The neck section is calculated by analyzing the electron density in a $(x,y)$ plane cutting the center of the neck. 
The right panel shows the 2D electron density distribution in that plane (blue dashed line). The red solid line 
represent the isocontour corresponding to a 0.00169~$e/\ang^{3}$ threshold density. 
The cross-section is defined as the area of the region limited by such isocontour. We can clearly see
that the neck cross-section is formed in this case by three Na atoms.}
\label{FigureS4}
\end{figure}

The neck formation during the retracting process opens the question of how to measure 
the neck cross-section. In particular, it is not obvious how to compare the
cross-sections of necks with similar structures.
Here we decided to use the distribution of the ground-state electron density as a mean to
measure the neck cross-section for an arbitrary structure. 
The density is computed in an $(x,y)$ plane passing through the middle of the junction
and the neck cross-section $A_{cs}$
is obtained as the area of the region where the electron density is larger than a given 
threshold value $\rho_{th}$ (see Fig.~\ref{FigureS4}).
In other words, we use the integral 
\begin{eqnarray}
A_{cs}=&\int\int dx dy \;\; f({\bf r})  &\nonumber \\
\mathrm{where} &  f({\bf r})=1 & \mathrm{if} \; |\rho({\bf r})| > \rho_{th}, \\
 \mathrm{and}     & f({\bf r})=0 & \mathrm{if} \; |\rho({\bf r})| < \rho_{th}.\nonumber 
 \label{e:cross_section}
\end{eqnarray}
The value of $\rho_{th}$ is arbitrary and was chosen here so
that the radius of an isolated Na atom is 2.88~$\ang$, a reasonable value
if we compare to Na bulk density (characterized by Wigner radius r$_s\sim$2.12~$\ang$)
and we take into account the spillage of charge towards vacuum in a finite object.
In any case, the specific value of the cross-section assigned to a particular neck structure is 
irrelevant (as far as {\it reasonable}), the importance of this method is the ability 
to continuously monitor the cross-section change as the structure evolves.

\section*{Split modes after nanocontact break}

\begin{figure}[htpb]
\centering
\includegraphics[width=8cm]{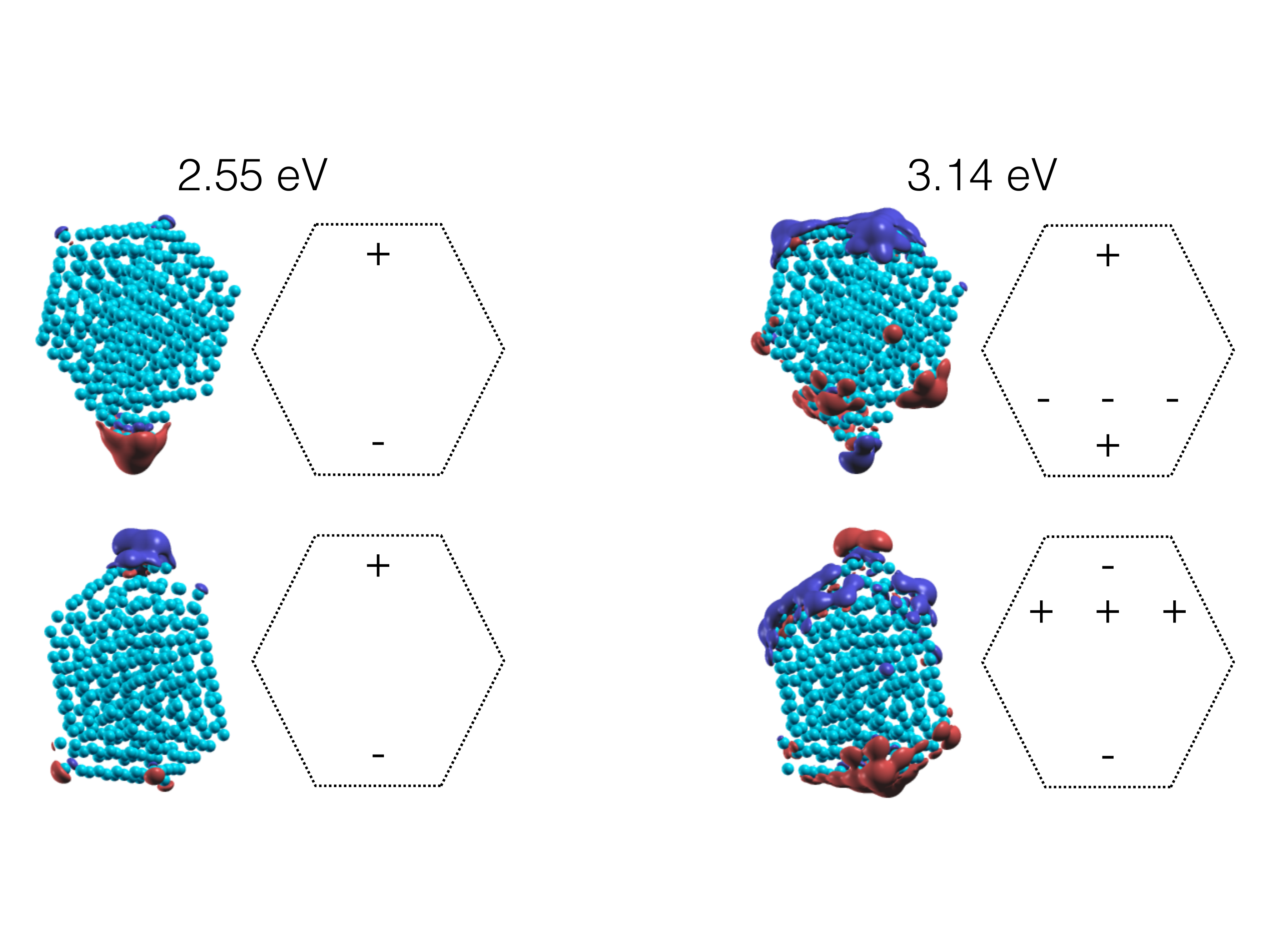}
\caption{Imaginary part of the induced charge density distribution for the frequencies of the
two plasmon resonances found after the break of the connecting neck and final separation 
of the clusters.}
\label{FigureS5}
\end{figure}

When the metal contact connecting the two clusters finally breaks and they separate, their final
structure differs considerably from the initial one.  The two clusters are not anymore identical 
and the mirror symmetry in the junction is lost. Furthermore, rather than in the
initial facet-to-facet configuration, the clusters now present some sort of tip-to-tip configuration.
With the rupture of the connecting neck, the CTP and CTP' modes that dominated
the optical response of the system during the whole retraction process disappear and give way to 
two modes, rather than a single BDP resonance.  
The new modes are found at slightly different energies.
Moreover, the  corresponding induced charge density distributions in Fig.~\ref{FigureS5} 
show different patterns. 
The lower energy resonance at 2.55~eV recalls a BDP mode, with the largest charge accumulations around the central gap.
The higher energy resonance at 3.14~eV has a more complex charge distribution, corresponding to higher order mode, showing charge accumulations both in the tips inside the cavity and in the facets of the clusters.

\section*{Real-time Dynamics: CTP Mode Across a Monatomic Chain}

\begin{figure*}[htpb]
\centering
\includegraphics[width=10cm]{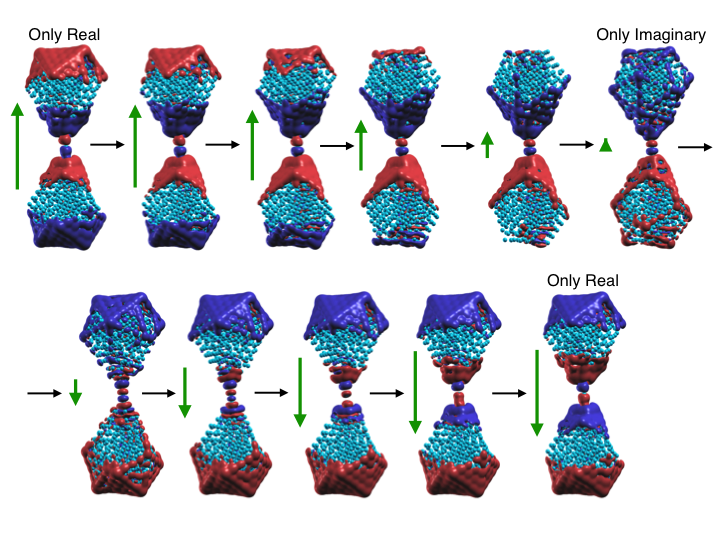}
\caption{Real time evolution of the induced density associated with the CTP mode when the cavity is connected
by a short monatomic wire. Black arrows indicate the time direction while green arrows indicate the direction 
and intensity of the external electric field in each instant. At $t=0$ and $t=\pi/\omega_{res}$ the distribution 
is completely derived from the real part $\delta n^{\prime}({\bf r},\omega_{res})$, at $t=\pi/2\omega_{res}$ it
is given by the imaginary part $\delta n^{\prime\prime}({\bf r},\omega_{res})$. }
\label{FigureS6}
\end{figure*}

The real space distributions of the induced charge density shown in the main text of the 
paper only correspond to the imaginary part. Therefore, although they 
can serve to characterize the resonant mode at a particular frequency, they may not
give a complete picture of the time evolution of the screening charge in the system. 

The imaginary part of the polarizability (induced density distribution) correspond to 
the optical absorption and provides information about the 
resonance frequencies (spatial distribution) of the different modes.
For a single, sharp excitation the maximum of the peak in the imaginary part of the 
polarizability coincides with a zero in the real part of the polarizability. Therefore,   when at resonance the
external field and the induced dipole in the system are always out of phase.
However, when dealing with a system having multiple resonances and especially at low frequencies,
even at resonance the real part 
can be different from zero and 
have a non-negligible intensity as compared to the imaginary part. 
In these cases, at the resonance frequency the real time evolution of the induced charge is a combination 
of the contributions coming from 
the imaginary and real part of the density change,
\begin{eqnarray}
\delta n({\bf r},t;\omega_{res})= &
\delta n'({\bf r},t;\omega_{res}) cos(\omega_{res}t) \nonumber \\
& + \delta n''({\bf r},t;\omega_{res}) sin(\omega_{res}t),
\label{e:tevo1}
\end{eqnarray}
being the external electric field ${\bf E} = {\bf E}_0 cos(\omega t)$.
The real part is in phase with the external field and the imaginary part representing the out-of-phase (resonant)
component of the total density change. 
This is the case for the CTP low energy mode seen in the polarizability once the mono-atomic neck is formed,
that shows a large real part of the induced density charge even at resonance. 

The sequence of imagines shown in Fig.~\ref{FigureS6} describes the evolution in time 
of the CTP mode across the cavity when there is a monatomic wire connecting the clusters.
At $t=\pi/2\omega_{res}$ we find the imaginary part of the induce charge, that characterizes
the mode resonant at that frequency. As expected, 
the dipole pattern extents all over the whole system with the presence with a node in the center of 
the junction. 
Thus, it corresponds to the expected charge transfer among the clusters. 
In contrast, the real part, found at $t=0$ and  $t=\pi/\omega_{res}$, presents 
a pattern formed by dipoles placed on each cluster that resembles a BDP mode. 

From a physical point of view what we see here is quite transparent. The monatomic wire across the
junction represents a bottleneck for electron conduction as compared to the facile movement 
of charges within each of the clusters. As a consequence, electrons can easily move across each 
of the clusters and react fast to the applied external field. However, when they reach the gap
in the center of the system they accumulate there, since the movement of charge across that gap
is limited by the monatomic chain, which provides just a single channel for electron conduction.

\section*{Conductance at Optical Frequencies}

\begin{figure}[htpb]
\includegraphics[width=7cm]{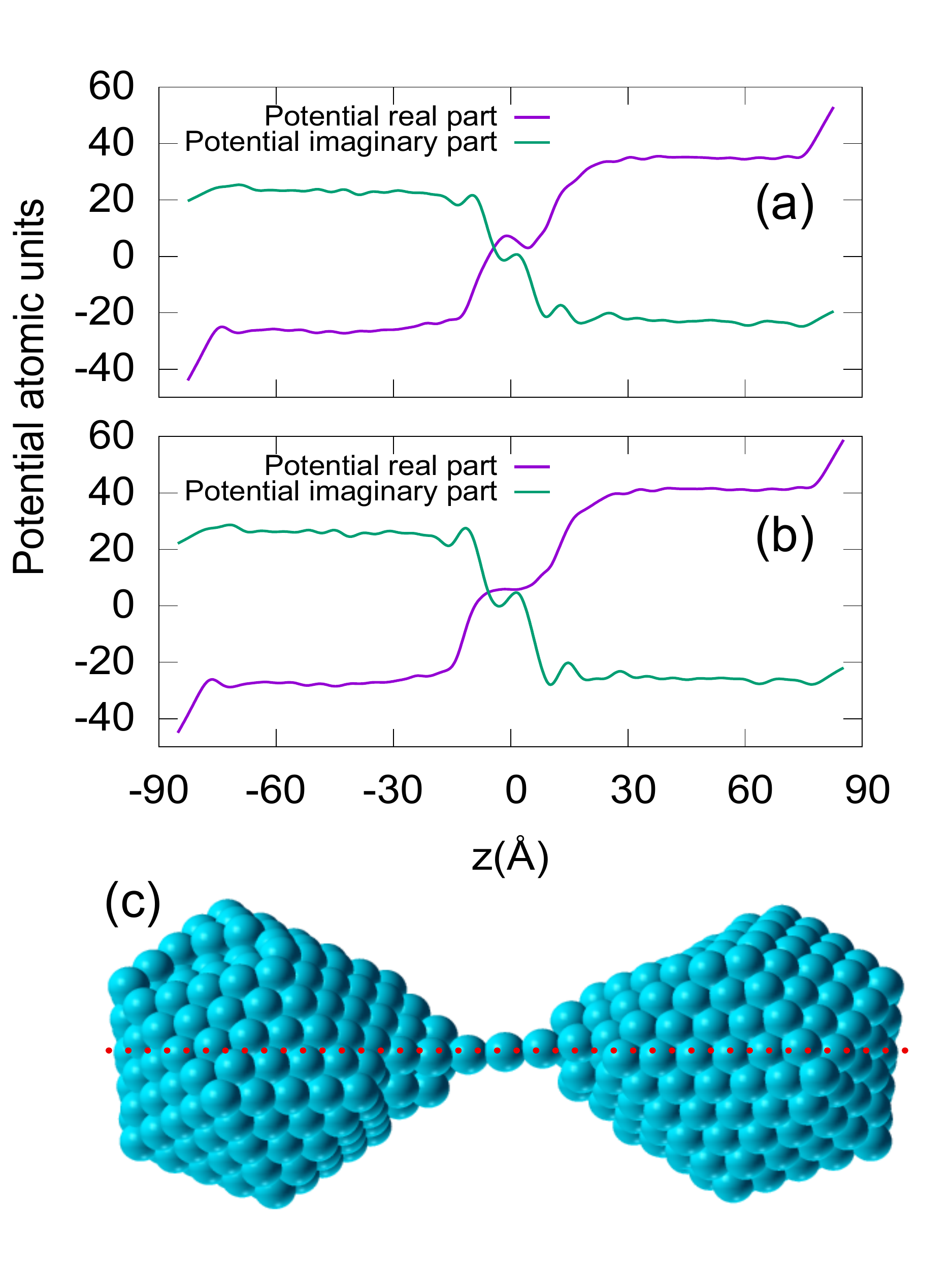}
\caption{In panels (a) and (b) we can find the real and imaginary part of the total electrostatic potential when a external 
 field of 1 a.u. is applied along the dimer axis. Panel (a) corresponds to a nominal distance of 31.3~$\ang$ and panel (b) for a distance of 34.1~$\ang$. The profiles are calculated along the line shown in panel (c) and for 
 the CTP resonance frequency. }
\label{FigureS7}
\end{figure}

The conductance is defined as the ratio 
between the current flowing across and the bias applied to a particular structure.
The conductance  
is frequently used to characterized the dc transport properties of nanowires. For wires of atom-sized cross-sections
the dc conductance can be quantized, i.e., it appears in multiples of the quantum of conductance G$_0$.
 In principle, it is also possible to calculate the neck conductance at optical frequencies from the 
 knowledge of the current passing through the neck, 
Eq.~\ref{e:I_max}, and the total potential drop between the two clusters. 
In this case, the potential in real space and frequency domain 
is a complex number calculated from the complex induced density charge 
distribution. The current also has a real and imaginary part. Thus, the result
in this case will be a complex number reflecting the capacitive component of the
impedance of the system.

In order to extract the potential drop across
the junction in a meaningful way it is necessary that the  
potential inside each cluster becomes constant.
This condition is met for configurations having a monatomic neck and at low frequencies. 
However, when the neck becomes wider and the energy of the mode increases is not possible anymore 
to define a potential drop restricted to the neck, i.e., the mismatch between 
the transmission through the neck and the clusters themselves is not sufficiently large that
the potential drop localizes mainly along the neck.
In particular, the imaginary part of the potential strongly varies inside the clusters.
Due to this behavior of the imaginary part of the potential profile
 it is not possible to define a voltage drop for wide contacts. Thus, we focus on configurations
 formed by clusters
 connected by monatomic chains. 
 
In Fig.~\ref{FigureS7}  (a) and (b) we depict 
the real and imaginary part of the total electrostatic potential for a nominal gap size of 31.3~$\ang$ and 34.1~$\ang$ 
plotted along the path in panel (c). As can be clearly seen, a bias drop across the monatomic chain
can be meaningfully defined for these geometries.
The computed conductance for these two cases are:
\begin{eqnarray}
G(\omega_{CTP})= &(4.52+ i2.21)\times10^{-5}~\mathrm{S}, \; & \; d=31.3~\ang  \nonumber \\ 
G(\omega_{CTP})= &(4.13+ i2.35)\times10^{-5}~\mathrm{S}, \; & \; d=34.1~\ang.  \nonumber
\end{eqnarray}
The corresponding moduli $|G(\omega_{CTP})|$ of the conductance are, respectively, 
 $5.03\mathrm{x}10^{-5}$ (0.65~G$_0$) for d=31.3~$\ang$  and $4.75\mathrm{x}10^{-5}$ (0.61~G$_0$) 
 for d=34.1~$\ang$, where G$_0$ = $7.75\mathrm{x}10^{-5}$~S is the quantum of conductance. 

\bibliography{Marchesin_etal}